\documentclass[aps,prd,twocolumn,superscriptaddress]{revtex4-1}

\usepackage{longtable}
\usepackage{grffile} 
\usepackage{graphicx} 
\usepackage{amsmath}    
\usepackage{hyperref}   
\usepackage{subfigure}  
\usepackage{color}
\usepackage{comment}

\def\sh2{\hat s^2}
\def\th2{\hat t^2}
\def\uh2{\hat u^2}

\begin{document}


\title{Investigating the broadening phenomenon in two-particle correlations induced by gluon saturation}

\author{K. Cassar}\email{kiera.cassar@stonybrook.edu}
\affiliation{School of Communication and Journalism, Stony Brook University, Stony Brook, NY 11794, U.S.A}

\author{Z. Wang}\email{wang.zhen@sdu.edu.cn}
\affiliation{Institute of Frontier and Interdisciplinary Science \& Key Laboratory of Particle Physics and Particle Irradiation (Ministry of Education), Shandong University, Qingdao, Shandong, 266237, China}

\author{X. Chu}\email{xchu@bnl.gov}
\affiliation{Physics Department, Brookhaven National Laboratory, Upton, NY 11973, U.S.A}

\author{E.C. Aschenauer}\email{elke@bnl.gov}
\affiliation{Physics Department, Brookhaven National Laboratory, Upton, NY 11973, U.S.A}

\begin{abstract}

It has been found that the gluon density inside the proton grows rapidly at small momentum fractions. Quantum Chromodynamics (QCD) predicts that this growth can be regulated by nonlinear effects, ultimately leading to gluon saturation. Within the color glass condensate framework, nonlinear QCD effects  are predicted to suppress and broaden back-to-back angular correlations in collisions involving heavy nuclei. While suppression has been observed in various experiments in $d/p$$+$A collisions compared to $p$$+$$p$ collisions, the predicted broadening remains unobserved. This study investigates the contributions of intrinsic transverse momentum ($k_T$), which is associated with saturation physics, as well as parton showers and transverse motion from fragmentation ($p_T^{\mathrm{frag}}$), which are not saturation dependent, to the width of the correlation function. Our findings show that the non-saturation dependent effects, especially the initial-state parton shower and $p_T^{\mathrm{frag}}$, which occur independently of the collision system, smear the back-to-back correlation more than gluon saturation does, making the broadening phenomenon difficult to observe.
\end{abstract}

\pacs{}
\maketitle

\section{Introduction}
An important focus of nuclear science is the origin and structure of the nucleus and the nucleons within it, which account for essentially all the mass of the visible universe. Half a century of investigations have revealed that nucleons themselves are composed of quarks, bound together by gluons, and have led to the development of the fundamental theory of Quantum Chromodynamics (QCD). Recent-generation colliders have provided ample data to extract the quark momentum distributions inside the nucleon, while a detailed exploration of the gluon dynamics is still limited~\cite{H1:2015ubc}. Particularly, our understanding of the so-called gluon saturation~\cite{Gribov:1984tu,Armesto:2004ud,Gelis:2010nm,Albacete:2010pg,Tuchin:2009nf,Kovchegov:2012mbw,ALBACETE20141,Morreale:2021pnn}, predicted by QCD to emerge in high energy collisions, remains incomplete. 

According to QCD, partons inside the nucleons are distributed as described by the parton distribution functions (PDFs), $f_{i}(x, Q^{2})$, parameterized by a fraction of the hadron’s longitudinal momentum fraction, $x$, carried by the parton that is being probed at the scale $Q^{2}$. It is well known that the concept of linear evolution on PDFs through the
Dokshitzer-Gribov-Lipatov-Altarelli-Parisi (DGLAP) formalism \cite{Gribov:1972ri,Dokshitzer:1977sg,Borah:2012ey} is a key ingredient for perturbative QCD (pQCD) and has been successfully
applied to understand many hard processes in high-energy collisions. 
Utilizing factorization theorems, PDFs can be extracted through global fits from, for example, cross section data~\cite{H1:2009pze}. A large number of high-energy physics experiments have established that the gluon distribution grows rapidly as $x$ decreases, due to gluon splitting. However, the gluon density cannot grow arbitrarily large, since this would violate the unitarity limit for forward scattering amplitudes at very high energies~\cite{Iancu:2003xm,Gelis:2010nm}. Recent experimental data~\cite{H1:2009pze} at small $x$ have provided us some intriguing evidence for the onset of a novel QCD regime, which cannot be fully described by linear QCD evolution approaches \cite{Gelis:2010nm}.

When the gluon density at low $x$ becomes so large that the gluons of a fixed transverse size
$\approx 1/Q^{2}$ begin to overlap, the QCD evolution dynamics essentially becomes nonlinear \cite{Gribov:1983ivg,Mueller:1985wy}. 
In this regime, QCD predicts that gluons can recombine in the dense medium. 
Although this phenomenon has yet to be confirmed experimentally, gluon splitting is expected to be balanced by gluon recombination \cite{Mueller:1985wy,Hwa:2004in}, thereby taming the rapid growth of the gluon density. In contrast to DGLAP, which describes how PDFs evolve with $Q^{2}$ at large 
$x$, different approaches have been developed to capture the nonlinear gluon dynamics at small $x$ under the color glass condensate (CGC) framework, such as the Balitsky-Kovchegov (BK) \cite{Balitsky:1995ub} and the Jalilian-Marian-Iancu-McLerran-Weigert-Leonidov-Kovner (JIMWLK) \cite{Iancu:2001ad} evolution equations.
Specifically, the BK equation models the evolution of the scattering amplitude for a color dipole (a quark-antiquark pair) interacting with a target hadron or nucleus as a function of $x$. Meanwhile, the JIMWLK equation governs the evolution of correlators of Wilson lines, which encode how a fast-moving quark or dipole interacts with the color field.
Quantitatively, the saturation scale $Q_{s}$ is introduced to characterize the transition between different regimes of parton density. For $Q^{2}>Q_{s}^{2}$, the target hadron and nucleus is usually treated as a dilute system,
whereas for $Q^{2}<Q_{s}^{2}$, the target becomes highly saturated, exhibiting a large parton density.

Back-to-back di-hadron azimuthal angle correlations at large pseudo-rapidities have been proposed to be one of the most sensitive probes for directly accessing the underlying gluon dynamics involved in hard scatterings~\cite{MARQUET200741,Zheng:2014vka}. Within the CGC framework~\cite{McLerran:1993ni,McLerran:1993ka,iancu2002colour}, quark and gluon
scatterings at forward angles (large pseudo-rapidities) will interact coherently with gluons at low-$x$ in the nucleus~\cite{Guzey:2004zp}. As a result, the probability of observing associated hadrons is expected to be suppressed, and an angular broadening of the back-to-back di-hadron correlation in $p(d)+$A collisions is predicted in comparison 
to $p$+$p$ collisions~\cite{Kharzeev:2004bw,Marquet:2007vb}.

From the experimental perspective, finding definitive evidence for nonlinear QCD effects is a primary objective of the
Relativistic Heavy Ion Collider (RHIC) Cold QCD program, the Large Hadron Collider (LHC) forward physics program, and the future electron ion collider (EIC). Previous results from collisions between hadronic systems, i.e., $p+$A or $d+$A at RHIC and the LHC provide a window into extracting the parton distributions of nuclei at small $x$. Several RHIC measurements~\cite{BRAHMS:2003sns,Arsene:2004ux,Adler:2004eh,PHENIX:2006feu,Adams:2006uz} have shown that the hadron yields at forward pseudo-rapidities (deuteron-going direction) are suppressed in $d+$Au collisions relative to $p$+$p$ collisions. 
Measurements by the LHCb experiment  \cite{LHCb:2017yua,LHCb:2023kqs} indicate the suppression of $D$ meson production in $p$+Pb collisions relative to $p$+$p$ collisions in the forward direction (proton-going direction).
For correlation measurements, on one hand, both RHIC \cite{Adare:2011sc,STAR:2021fgw} and the LHC \cite{ATLAS:2019jgo} observed this predicted suppression in nucleus-involved collisions compared to the baseline $p$+$p$ collisions at different collision energies, suggesting hints of gluon saturation. On the other hand, none of the experiments has observed the broadening phenomenon predicted by CGC to date.

\section{Method}
The puzzle of the unobserved broadening phenomena has motivated and was partially explained by several theoretical and phenomenological studies~\cite{Zheng:2014vka,Hanninen:2023wbp,Mantysaari:2022kdm,Caucal:2023fsf}. Reference \cite{Zheng:2014vka} reviewed the feasibility of measuring two charged hadron correlations in simulated $e$+A and $e$+$p$ collisions at the future EIC to explore its capability to study saturation. Gluon saturation was implemented using unintegrated gluon distributions (UGDs) in $e$+A collisions, with an intrinsic transverse momentum ($k_{T}$) assigned to gluons in the initial-state. 
This transverse motion of partons inside hadrons can be effectively included by assuming that $k_{T}$ follows a Gaussian distribution.
Reference~\cite{Lappi:2013zma} discusses in detail how to estimate $k_{T}$ for $p$$+$$p$ and $p$$+$Pb collisions at various collision energies, postulating the saturation scale $Q_{s}$ being strongly correlated to the value of the $k_{T}$. In the phase space where STAR forward di-$\pi^{0}$ data can probe, we assigned $k_{T}$ to be 0.5 and 1.0 GeV/$c$ for $p$$+$$p$ and $p$$+$Au collisions, respectively.
A common practice is to compare the correlation functions with a Gaussian width of $k_{T}$ = 1.0 GeV/$c$ in $p+p$ collisions to those with a width of $k_{T}$ = 0.5 GeV/$c$, and evaluate whether the back-to-back correlation function is broadened. 

It is important to note that the CGC calculations have entered the next-leading-order (NLO) era \cite{Hanninen:2023wbp,Mantysaari:2022kdm,Caucal:2023fsf}, as it has been observed that the leading-order (LO) calculations are no longer able to describe the global data for two-particle correlation measurements, especially their width. 
NLO contributions, such as parton showers, broaden the away-side peak of the di-hadron
correlation function, just as saturation is predicted to do. The contribution from parton showers is effectively included into the Sudakov factor. More importantly, this contribution exists in all types of collision systems no matter what species the beam is. However,
it remains unclear how the parton shower effect might be modified in the nuclear medium. Without a detailed understanding, it is difficult to draw any definite conclusions about the saturation effects, as parton showers and saturation effects are always entangled.

One final contribution, the transverse momentum induced by the fragmentation process, $p_{T}^{\mathrm{frag}}$, can also modify the shape of the back-to-back correlation. In pQCD calculations within the collinear factorization framework, the PDFs and fragmentation functions do not contain any transverse momentum dependence. Therefore, the transverse momentum of hadrons produced in
the final state is given by $p_{T} = z\hat{p_{T}}$, where $\hat{p_{T}}$ and $p_{T}$
are the transverse momentum of the parton and hadron, respectively. Here, $z$ represents the momentum fraction of a hadron with respect to its mother parton. This relation should be revised if one incorporates transverse momentum effects into fragmentation functions. Like the intrinsic $k_{T}$, $p_{T}^{\mathrm{frag}}$ can be effectively modeled by assuming that it follows a Gaussian distribution.
A wider back-to-back correlation function, compared to the case with $p_{T}^{\mathrm{frag}}$ = 0 GeV/$c$, is expected to be generated by setting the Gaussian width of $p_{T}^{\mathrm{frag}}$ to be 0.4 GeV/$c$. This value is extracted from a global analysis~\cite{Anselmino:2013lza} using data from the EMC~\cite{EuropeanMuon:1991sne}, HERMES~\cite{HERMES:2009uge,HERMES:2012uyd}, and COMPASS~\cite{COMPASS:2013bfs} experiments.
A comparison of the generated correlation function is compared to the output from the collisions without assigning $p_{T}^{\mathrm{frag}}$. 

In this paper, we perform a detailed study of the effects from intrinsic $k_{T}$, parton showers, and $p_{T}^{\mathrm{frag}}$ to the correlation function. The intrinsic $k_{T}$ is assigned following gluon saturation models, while parton showers and $p_{T}^{\mathrm{frag}}$ exist independent of saturation physics in all types of collision systems. 

\begin{figure*}[t]
\begin{center}
\includegraphics[width=0.47\textwidth]{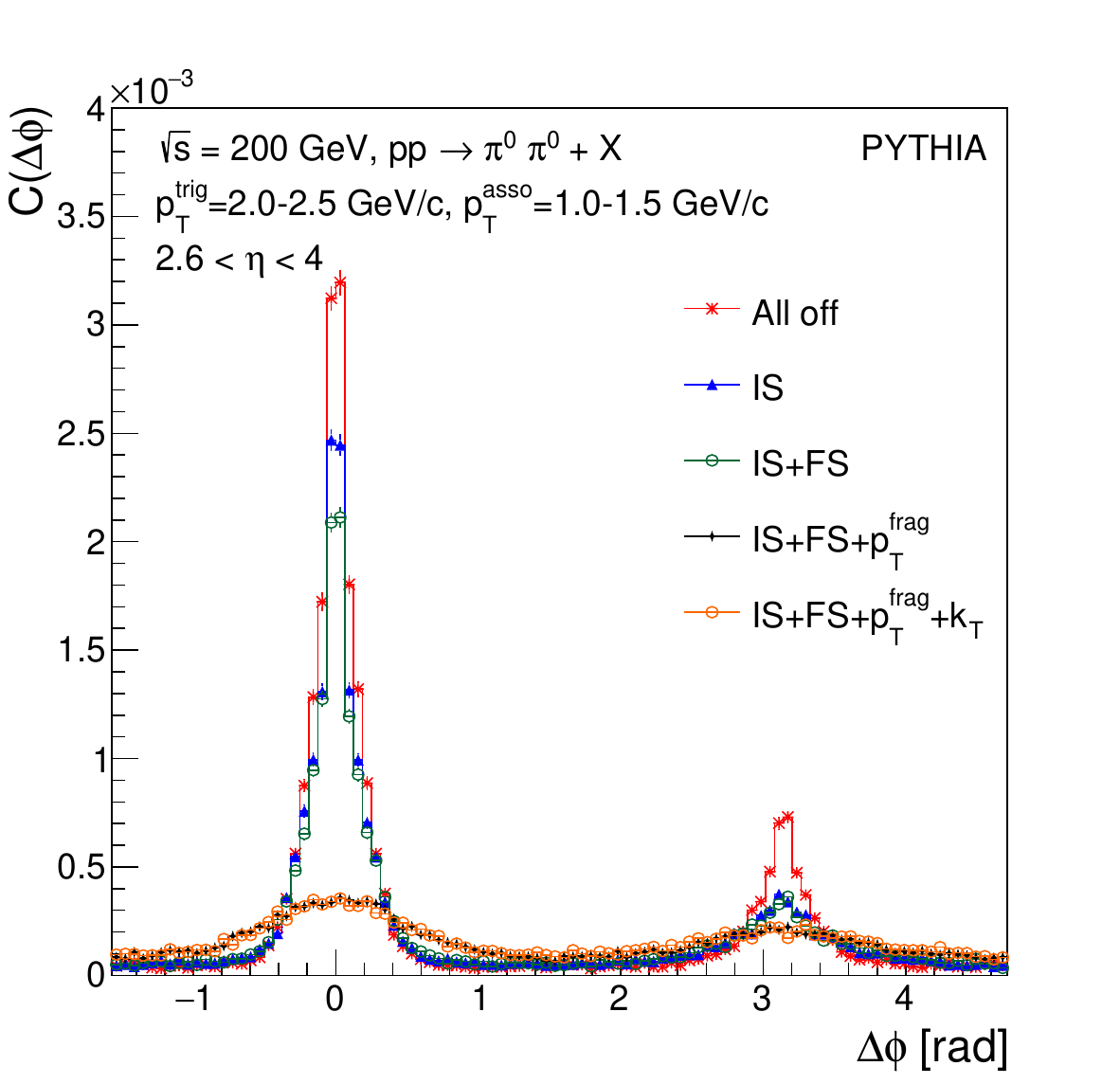}
\includegraphics[width=0.47\textwidth]{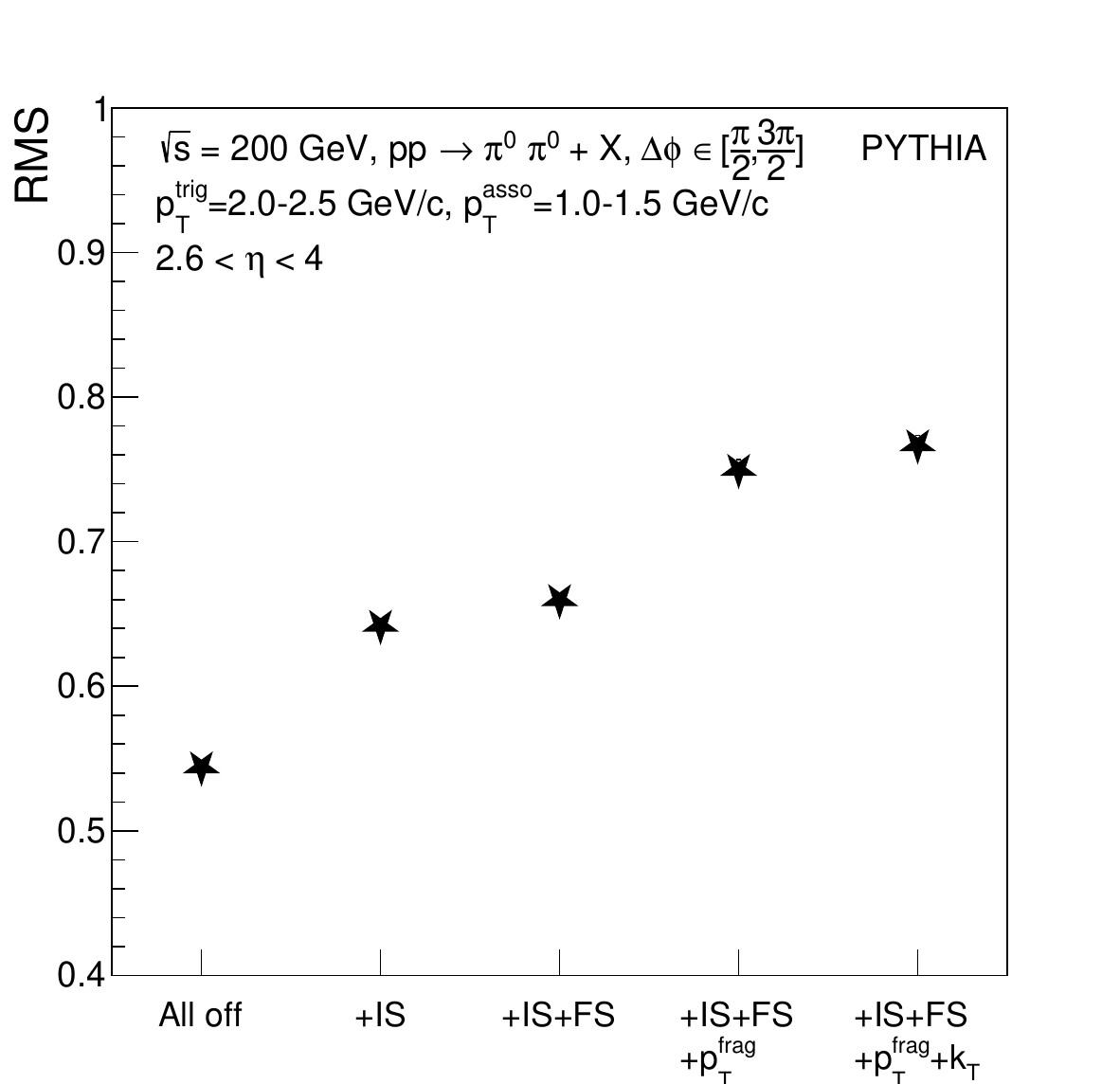}
\end{center} 
\caption{Left: The correlation function of forward di-$\pi^{0
}$ production for different intrinsic $k_{T}$ assumptions, parton showers, and $p_T^{\textrm{frag}}$. Right: The RMS of the away-side peak for different assumptions. The Gaussian width of $k_{T}$ and $p_{T}^{\mathrm{frag}}$ are 0.5 and 0.4 GeV/$c$, respectively. 
The simulated $p$+$p$ data are for di-$\pi^{0
}$s with a center-of-mass energy of 200 GeV 
with $2.0 < p_{T}^{\mathrm{trig}} < 2.5~\mathrm{GeV/}c$ and $1.0 < p_{T}^{\mathrm{asso}} < 1.5~\mathrm{GeV/}c$, at $2.6 < \eta < 4$.}
\label{fig:f1}
\end{figure*}

\section{Monte Carlo Set Up}
The results of this paper are based on simulations using the PYTHIA-$6$ Monte Carlo (MC) generator~\cite{Sjostrand:2006za} with the standard STAR tune~\cite{STAR:2019yqm}, employing PDFs from the LHAPDF library~\cite{Bourilkov:2006cj} and JETSET for fragmentation processes. The factorization scale $\mu^{2}$ of $2\rightarrow 2$ processes is expressed as $\mu^{2} = \hat{p_{T}}^{2} + \frac{1}{2}Q^{2}$. We simulated 200 GeV $p$+$p$ collisions using CTEQ6L \cite{Pumplin:2002vw} as the input PDF for the two proton beams. 

\section{Simulation results and discussions}
In this paper, the observable is the correlation function of $\pi^{0}$ pairs (di-$\pi^{0}$) at forward pseudo-rapidity (2.6$<\eta<$4) in 200 GeV $p$+$p$ collisions.
In each pair, the trigger hadron is the one with the higher $p_{T}$ value, $p_{T}^{\mathrm{trig}}$, and the associated hadron is the one with the lower $p_{T}$ value, $p_{T}^{\mathrm{asso}}$.
The correlation function, $C(\Delta\phi)$, is defined as
$C(\Delta\phi) = \frac{N_\mathrm{pair}(\Delta\phi)}{N_\mathrm{trig}\times \Delta \phi _\mathrm{bin}}$, 
where $N_\mathrm{pair}$ is the yield of correlated
trigger and associated hadron pairs, $N_{\mathrm{trig}}$ is the
trigger hadron yield, $\Delta\phi$ is the azimuthal angle difference between the trigger hadron and associated hadron,
and $\Delta\phi_{\mathrm{bin}}$ is the bin width of the $\Delta\phi$ distribution.
The correlation function is fitted over the range $\Delta\phi = -\pi/2$ to $\Delta\phi = 3\pi/2$ using two individual Gaussians at the near- ($\Delta\phi=$ 0) and away-side ($\Delta\phi=\pi$) peaks, together with a constant term representing the pedestal. 
The pedestal originates from two $\pi^{0}$s that are not correlated with each other.
The area of the away-side peak is the integral of the correlation function over $\Delta\phi = \pi/2$ to $\Delta\phi = 3\pi/2$ after pedestal subtraction, describing the back-to-back $\pi^0$ yields per trigger particle. 
The corresponding width is defined as the standard deviation ($\sigma$) of the away-side peak according to the fit. 

The parton momentum fraction $x$ in the nucleon is proportional to the transverse momentum of the two hadrons, and $Q$ can be approximated as their average $p_{T}$. By varying $p_{T}$, one can effectively probe different values of $x$ and $Q^{2}$.
At low $p_{T}$, the gluon density is expected to be large and potentially in a saturation regime. When $p_{T}$ is high, $x$ and $Q^{2}$ are not small enough to reach this nonlinear regime. 
Recent STAR results on forward di-$\pi^{0}$ correlations~\cite{STAR:2021fgw} suggest that the transverse momentum range ($p_{T}^{\mathrm{trig}}$ = 2$-$2.5 GeV/$c$ and $p^{\mathrm{asso}}_T$ = 1$-$1.5 GeV/$c$) is sufficiently small to observe the signatures of nonlinear effects. In this study, the correlation functions concentrate on this $p_{T}$ combination, where the probed $x_{2}$ covers a wide range from $10^{-4}$ to $\sim0.5$ and the mean value of $Q^{2}$ is 2.5 GeV$^{2}$.

\begin{figure}[t]
\begin{center}
\includegraphics[width=0.47\textwidth]{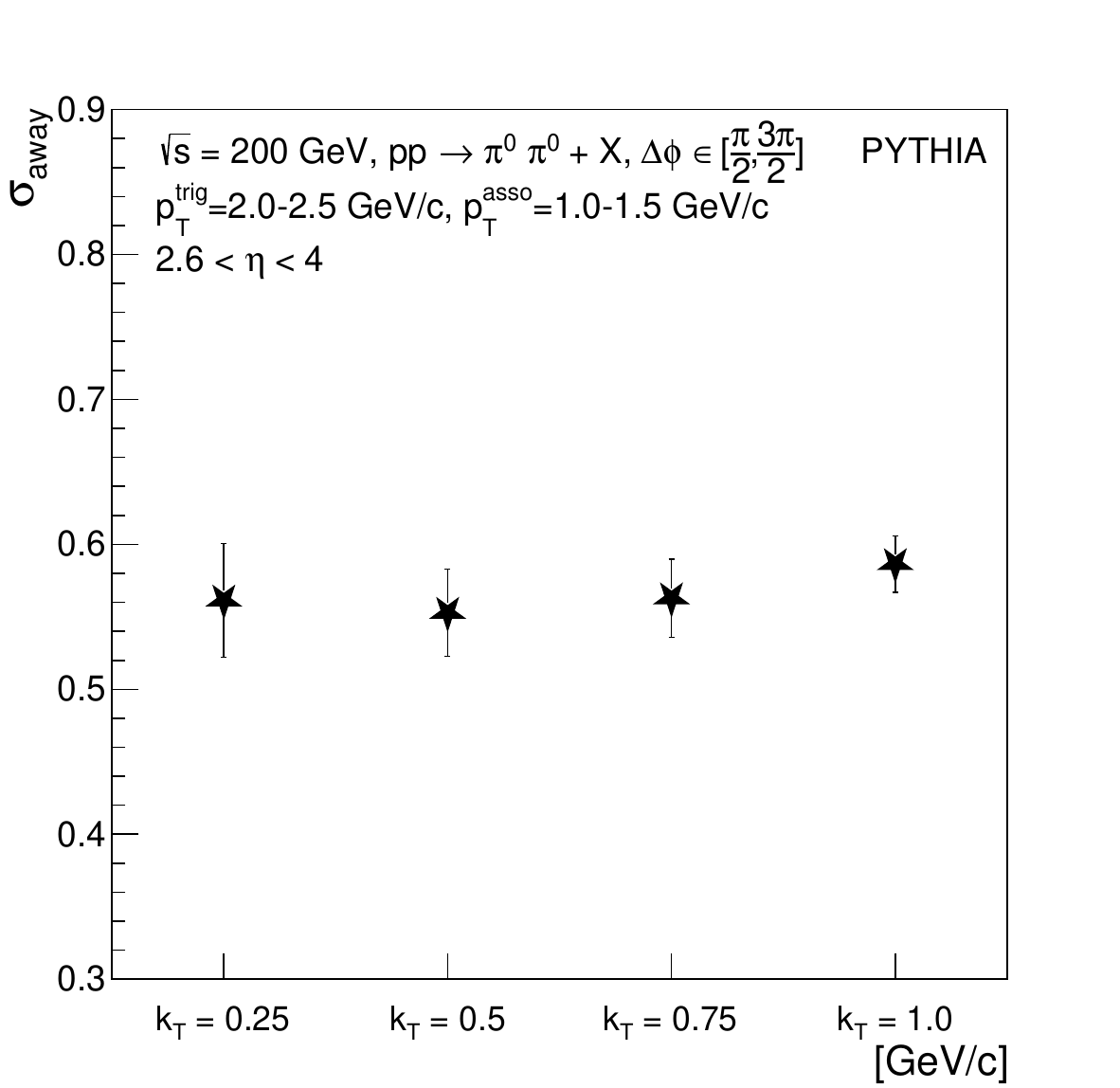}
\end{center} 
\caption{ 
The extracted width of the away-side peak at different intrinsic $k_{T}$. The simulated $p$+$p$ data are for di-$\pi^{0
}$s with a center-of-mass energy of 200 GeV 
with $2.0 < p_{T}^{\mathrm{trig}} < 2.5~\mathrm{GeV/}c$ and $1.0 < p_{T}^{\mathrm{asso}} < 1.5~\mathrm{GeV/}c$, at $2.6 < \eta < 4$.}
\label{fig:f2}
\end{figure}

The left panel of Fig.~\ref{fig:f1} shows the impact of sequentially adding all available effects into the MC simulation of the azimuthal correlation function.
The red points illustrate the di-$\pi^{0}$ correlation with all the effects off, serving as a baseline correlation result. As expected, the
correlation function is strongly peaked at $\Delta \phi =0$ and $\pi$, for this setting.
When only initial-state parton shower (IS) contribution is included (blue points), the correlation function is significantly suppressed at both peaks. Adding final-state parton shower (FS) contribution (green points) leads to a reduced near-side correlation, while the away-side peak remains almost unchanged. Once transverse motion from fragmentation $p_T^{\textrm{frag}}$ is introduced (black points), both the near- and away-side peaks are broadened significantly. 
In the right panel of Fig.~\ref{fig:f1}, we quantify the broadening phenomena from each source by presenting the RMS width of the away-side peak. 
The largest contributions to the broadening come from IS and $p_T^{\textrm{frag}}$.
We conclude, the initial-state parton shower and fragmentation process, independent of gluon saturation, are the major effects modifying the width of the away-side correlation function at RHIC kinematics. Specifically, it is expected the scattering at RHIC is characterized by lower transverse momentum scales, which makes the fragmentation process more influential compared to the much higher energy collisions at the LHC. In other words, $p_T^{\textrm{frag}}$ must be considered when predicting the correlation function to study the saturation physics.  

To further explore potential saturation effects, we study the impact of intrinsic $k_{T}$ by varying its Gaussian width in simulations. With parton showers and $p_T^{\textrm{frag}}$ enabled, the correlation functions are measured while varying the values of $k_{T}$, which are selected to be 0.25, 0.5, 0.75, and 1.0 GeV/$c$. The selections of 0.5 and 1.0 GeV/$c$ are motivated by previous studies~\cite{Lappi:2013zma}, that at RHIC energies where the average $x$ probed is $10^{-2}$, $Q_{s}$ is 0.5 and 1.0 GeV/$c$ for proton and gold nuclei, respectively. As seen in Fig. \ref{fig:f2}, the extracted width of the away-side peak ($\sigma_{\mathrm{away}}$) remains approximately constant with increasing $k_{T}$. The error bars indicate the statistical uncertainties obtained from the fits. Figure \ref{fig:f2} shows that once NLO contributions such as parton showers and fragmentation effects are included, the width of the back-to-back correlation function is insensitive to variations in intrinsic $k_{T}$. This is a possible explanation of the observed contradiction between different theoretical predictions and the data in experiments across different colliders, with respect not observing a broadening despite being predicted.

\section{summary}
The modification of back-to-back di-hadron correlation functions at forward pseudo-rapidities has been investigated, exploring the effects of gluon saturation as well as other factors, including parton showers and the fragmentation process. Gluon saturation was effectively implemented by assigning an intrinsic transverse momentum to the partons in the initial state. It has been shown that all of these factors can create notable changes in the shape of the back-to-back correlation function in 200 GeV $p$+$p$ collisions. Among all of them, the initial-state parton shower and fragmentation $p_{T}$ play the major role. 
As a result, it is difficult to observe the broadening phenomenon of two-particle correlation measurements predicted as part of gluon saturation models. We conclude that the inclusion of the NLO contributions are essential for inclusion in CGC calculations of back-to-back di-hadron correlation functions

\begin{acknowledgments}
This project was supported in part by the U.S. Department of Energy, Office of Science, Office of Workforce Development for Teachers and
Scientists (WDTS) under the Science Undergraduate Laboratory Internships Program (SULI), the Brookhaven National Laboratory (BNL) Physics Department under the BNL Supplemental
Undergraduate Research Program (SURP), the Laboratory Directed Research and Development (LDRD) 24-071 subtask and
LDRD 25-029 projects, and the BNL Cold QCD Group funding. 
\end{acknowledgments}

\bibliography{reference.bib}
\end{document}